# A Study of CAPTCHAs for Securing Web Services

M. Tariq Banday[1], N. A. Shah[2]

[1] P. G. Department of Electronics and Instrumentation Technology,
University of Kashmir, Srinagar - 6, India
*Email: sgrmtb@kashmiruniversity.ac.in*

[2] P. G. Department of Electronics and Instrumentation Technology,
University of Kashmir, Srinagar - 6, India
*Email: nassgr@yahoo.com*

**Abstract:** *Atomizing various Web activities by replacing human to human interactions on the Internet has been made indispensable due to its enormous growth. However, bots also known as Web-bots which have a malicious intend and pretending to be humans pose a severe threat to various services on the Internet that implicitly assume a human interaction. Accordingly, Web service providers before allowing access to such services use various Human Interaction Proof's (HIPs) to authenticate that the user is a human and not a bot. Completely Automated Public Turing test to tell Computers and Humans Apart (CAPTCHA) is a class of HIPs tests and are based on Artificial Intelligence. These tests are easier for humans to qualify and tough for bots to simulate. Several Web services use CAPTCHAs as a defensive mechanism against automated Web-bots. In this paper, we review the existing CAPTCHA schemes that have been proposed or are being used to protect various Web services. We classify them in groups and compare them with each other in terms of security and usability. We present general method used to generate and break text-based and image-based CAPTCHAs. Further, we discuss various security and usability issues in CAPTCHA design and provide guidelines for improving their robustness and usability.*

## 1. Introduction

HIPs [1] are schemes that allow a computer to distinguish a specific class of humans over a network. HIPs can be designed to distinguish a human from a computer, one class of humans from another or one particular human from another human. To do this, the computer presents a challenge that must be easy for that class of humans to pass, yet hard for non-members to pass. Additionally, the results must be verifiable by a computer, and the protocol must be publicly available [1]. CAPTCHA [2] is a class of HIPs that have been able to effectively prevent Web-bots from getting access to the Web services. CAPTCHA is a reverse Turing test based on text, image or audio based challenge response system. Various implicitly human interactions assumed services on the Internet use CAPTCHA techniques to ensure that the user of these services is a human and not a Web-bot. Web services and applications that use CAPTCHA methods for as HIP include chat rooms, search engines, password systems, online polls, e-mail services for account registrations, prevention of sending and receiving spam, blogs, messaging services, free content downloading services and detecting phishing attacks [3]. CAPTCHAs have been able to prevent the abuse of several Web services and thus offer advantages but at the same time its use poses various disadvantages. Text and image-based CAPTCHAs are designed hard and as such are unfriendly particularly for disabled and visually impaired people. Audio CAPTCHAs which are used as HIP for visually impaired people are very difficult to pass. CAPTCHAs increase load on servers because of requirement for image database and huge server processing and thus result in delay of Web page downloads and their subsequent refreshes. Further, CAPTCHAs pose an annoyance to genuine user. A good CAPTCHA minimizes these disadvantages by generating a CAPTCHA test that satisfies its various desired properties. These properties include i) automatic generation of the test, ii) quick and easy answer to the test, iii) acceptance to all humans or a class of humans, and iv) resistance to attacks with publically known protocol [2].

This paper studies various aspects of CAPTCHA methods that include its types, generation methods, robustness against attacks and various usability aspects. It presents relative merits and demerits of text and image based CAPTCHA methods. Section 2 presents a review of existing CAPTCHA schemes. Section 3 illustrates working of CAPTCHAs and general methods used for their generation. Section 4 discusses security and usability issues of CAPTCHA methods. It provides guidelines to improve security control of CAPTCHA methods against various possible attacks and guidelines to improve their usability. Finally in section 5, we conclude and present future research directions.

---

## 2. Types of CAPTCHA methods

### 2.1. Text-Based CAPTCHAs

CAPTCHA was initially devised by Andrei Broder and his colleagues in 1997 and in the same year Altavista website used this method as a HIP [2] in the same year. This method used a distorted English word that a user was asked to type. The distorted word was easier for users to understand but difficult for bots to recognize using OCR techniques. Text based CAPTCHAs are in the form of an image containing a difficult to recognize text string to be identified and typed by the user in a text box provided near the CAPTCHA image on the Web page. The CAPTCHA image is of low quality with different forms of noise and strong degradation applied to it.

Blum and Von Ahn in Collaboration with Yahoo devised EZ-Gimpy and Gimpy CAPTCHA [4] to protect chat rooms from spammers. These CAPTCHAs challenges have been broken by dictionary attacks which contained a limited number of words in them [5]. A more secure type of text based HIP, called reCAPTCHA [6] has been proposed by the same authors. Baffle Text CAPTCHA [7] is the Xerox Pato Alto Research Center (PARC) version of Gimpy test. Prominent text based CAPTCHA techniques include Scatter Type [8], Handwritten Word based CAPTCHA [9] and Human Visual System masking Characteristic CAPTCHA [10]. Various service providers on the Internet like PayPal, Hotmail and YouTube use their own versions of text based CAPTCHAs on their websites and update them with newer versions frequently. With an aim of improving usability of text-based CAPTCHAs, Richard Chow et al [11] have proposed a generic technique for converting a textual CAPTCHA into a clickable CAPTCHA. It proposes placement of multiple text CAPTCHA images in a grid among which some are English words while others are not. The user must click on all valid English words to pass this CAPTCHA test. Samples of various text-based CAPTCHAs are shown in figure 1.

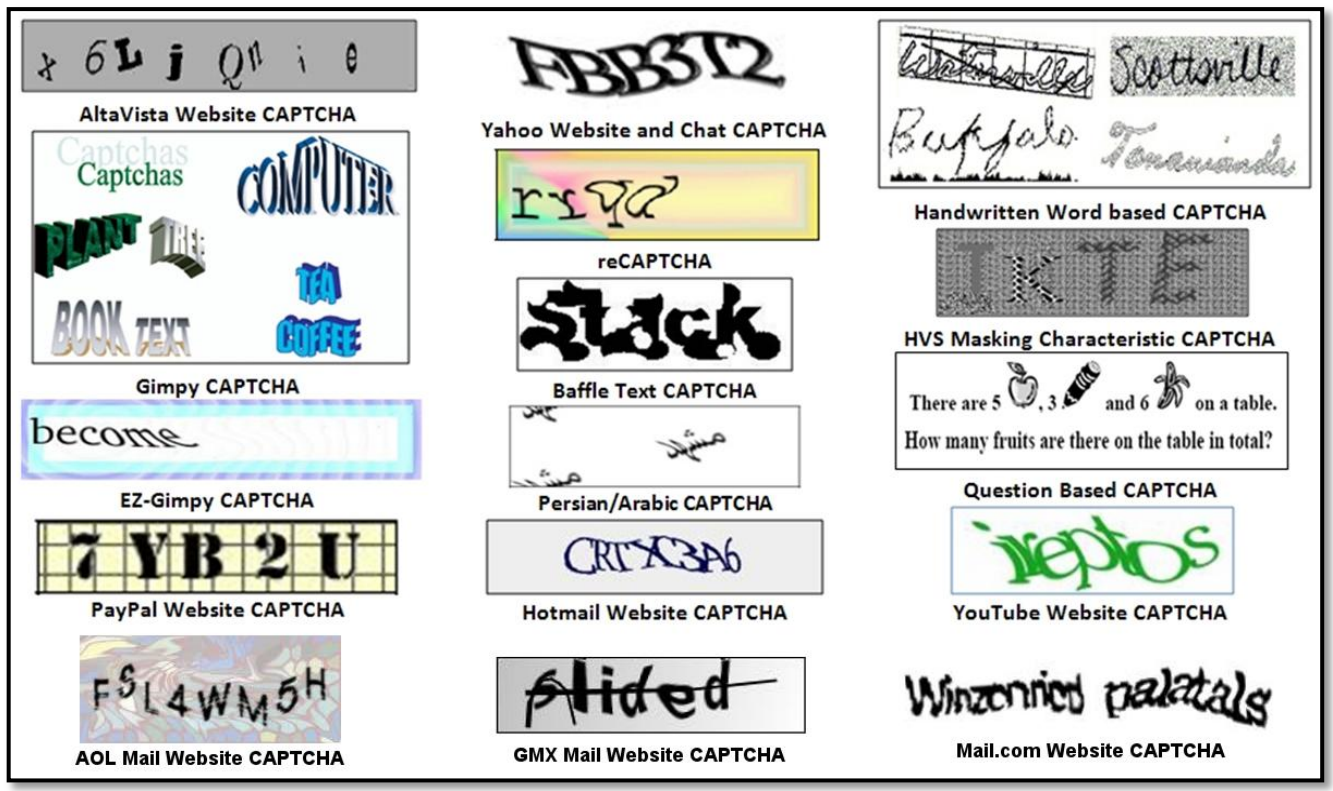

**Figure** 1: Samples of Text-based CAPTCHAs

### 2.2. Image-Based CAPTCHAs

Blum and Von Ahn proposed initially image-based CAPTCHA called ESP-PIX CAPTCHA [2]. It used a larger database of photographs and animated images of everyday objects. The CAPTCHA system presented a user with a set of images all associated with the same object or concept. The user was required to enter the object or concept to which all the images belonged to e.g. the program might present pictures of Globe, Volleyball, Planet and baseball expecting the user to correctly associate all these pictures with the word ball. In general image based CAPTCHAs present a visual pattern or concept that the user needs to identify and act accordingly. Different image-based CAPTCHA scheme use different patterns or concepts which are easy to be recognized by the users and difficult for the bot programs to simulate.

Besides several others, this class of CAPTCHA methods include: Microsoft Asirra [12], IMAge Generation for INternet AuthenticaTION (IMAGINATION) [13], Image Block Exchange [14] and Face Recognition [15] CAPTCHAs. Mosaic-based Human Interactive Proof called MosaHIP [16] proposes a CAPTCHA scheme for securing the download of resource against Web-bots. It uses a single larger image called mosaic image which is composed of smaller and partially overlapping real and fake pictures. The user needs to drag a resource expressed in form of movable text object on the web page and drop it onto the area of the mosaic picture containing the image indicated in the CAPTCHA image. Google has proposed a CAPTCHA method in which a user has to adjust randomly rotated images to their upright orientation [17]. Recently, an Image Flip CAPTCHA [18] method proposes use of a composite CAPTCHA image comprising of flipped and non-flipped images. The user needs to click on all images that appear as normal and without any flip applied to them. Samples of various image-based CAPTCHAs are shown in figure 2.

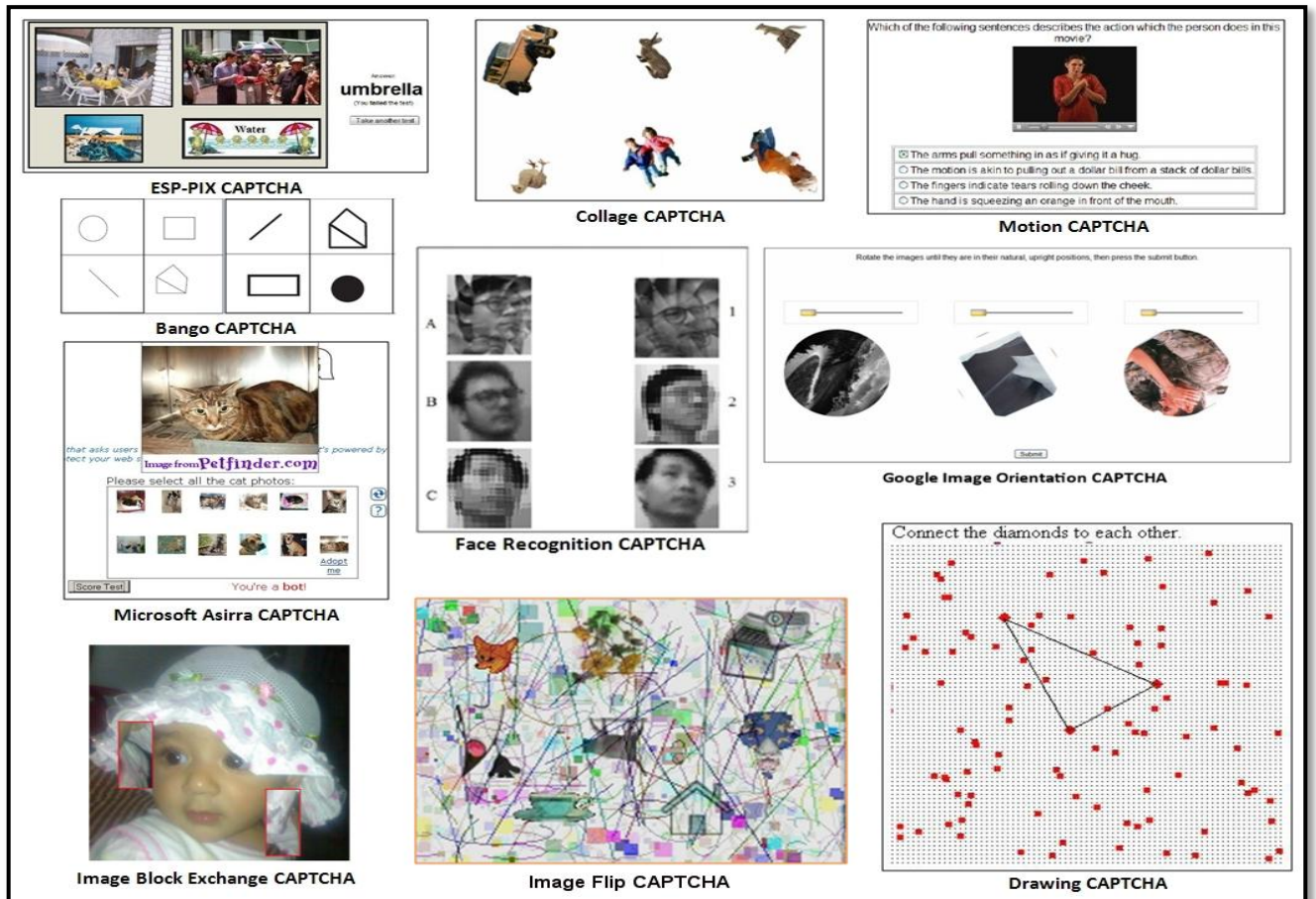

**Figure** 2: Samples of Image-based CAPTCHAs

### 2.3. Audio-Based CAPTCHAs

The first audio based CAPTCHA was implemented by Nancy Chan to provide an alternative to text based CAPTCHAs for visually impaired people. Audio CAPTCHAs [19] take a random sequence drawn from recordings of simple words or numbers, combine them and add some disturbance and noise to it. This recording is played when the user clicks a button provided on the web page. The CAPTCHA system then asks the user to enter the words and/or numbers in the recording. Audio CAPTCHAs are more difficult to solve, hard to internationalize and more demanding in terms of time and efforts in comparison to text and image CAPTCHAs. However, audio based CAPTCHA tests have become an alternative for visually impaired people. Most Web services include it in addition to text and image CAPTCHAs.

### 2.4. Other CAPTCHAs

Besides above types of CAPTCHA tests, Collaborative Filtering [20] and Implicit [21] CAPTCHA challenge have been proposed in literature. Collaborative filtering CAPTCHAs approaches differ from others in the scenes that CAPTCHA designers do not initially know the correct answer for their designed CAPTCHA, but measure it from human opinion. Implicit CAPTCHA proposes single click challenges distinguished as necessary browsing links which can be answered through experience of the context of the particular Website. A review of the existing CAPTCHA techniques is provided in [3, 22].

## 3. Working of CAPTCHA

A Web server may be holding both public and protected resources that may be in the form of web pages, data stored in a database or files or some other service intended to be used by human users on the client. User request for a resource is send by the client computer to the server, which is granted to it if the resource is not protected. In case the resource is CAPTCHA protected, the access is granted to it only after passing CAPTCHA test as depicted in figure 3.

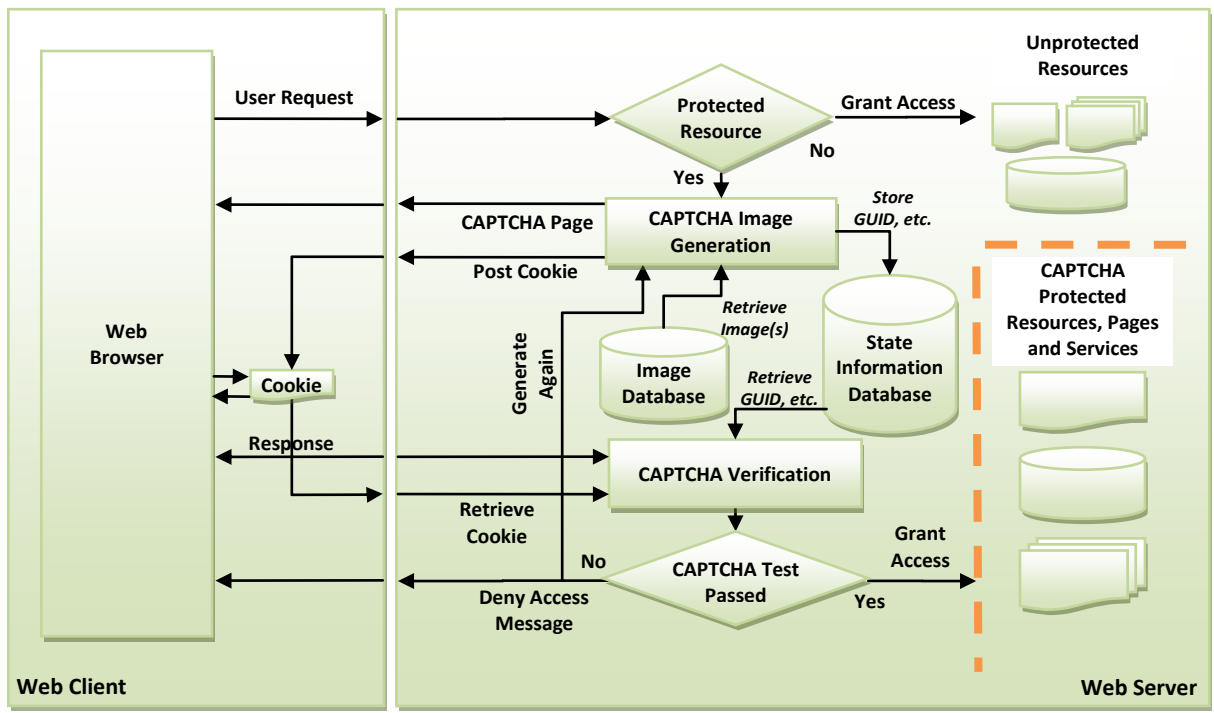

**Figure** 3: Working of CAPTCHA

The server uses some CAPTCHA image generation algorithm to generate a CAPTCHA image. Different CAPTCHA techniques use different algorithms for image generation which may employ use of images stored in an image database. The state information along with Global Unique Identifier (GUID) of the client and the CAPTCHA solution is stored in the State Information Database (SID) at the server. Storing GUID of the client ensures that only client that received CAPTCHA can produce a valid solution. Instead of storing the CAPTCHA solution and other state information on server in SID, it be may stored in hashed or encrypted form in a cookie on the client. A web page containing the generated CAPTCHA image and the cookie is posted to the client which renders it in a web browser to the user. A human operator responds to CAPTCHA test and the response is passed by the client to the server. The server verifies the authenticity of CAPTCHA solution by comparing the stored GUID and the GUID of the client sending the solution. The solution provided by the client is next compared with the solution stored in SID or cookie and accordingly either access is granted or denied. In case access is denied, a message is posted to the client and the process starts afresh. A CAPTCHA implementation may temporarily block access for a client if it repeatedly fails to respond to a number of CAPTCHA tests. Further, for a particular session once a CAPTCHA challenge has been passed by a client, subsequent accesses to protected resources on the server may be granted to it without putting it to further tests.

### 3.1. CAPTCHA Image Generation Process

It is not possible to generalize the algorithm for generation of CAPTCHA image; however, the steps listed below provide a guideline for creation of a basic text-based CAPTCHA image:

i. *Create a CAPTCHA image of desired dimension sufficient to hold the text string.*
ii. *Set the background color for the CAPTCHA image. Instead of choosing a solid background color, a pattern or a stored background image may be used. Some CAPTCHA tests use a simple* *white background while others fill the image with some form of noise.*

iii. *Generate n random characters from the designated character set and/or digit set. Generally upper and lower case English alphabets are used as character set and 10 Arabic numerals are used as digit set. A CAPTCHA test which is case insensitive and uses both alphabets and numbers thus has total combinations of 36 characters while as the one which is case sensitive, has a total combination of 62 characters. The character generation algorithm is made to generate those characters which are similar to one another in some manner and will make the string complex to be understood by the OCR techniques.*

iv. *Chose the font, font size, font style, font color and other related attributes. Most CAPTCHA implementations make use of personalized fonts and apply twist to characters so as to make OCR techniques to fail. Some CAPTCHA tests use different fonts and styles for different characters to make the test more secure.*

v. *Select a random spacing between each generated character. The spacing is chosen in a manner that some characters partially overlap one another. The text string is then placed on the CAPTCHA image.*

vi. *Optionally, generate lines and arcs or other objects with desired parameters and place it on the CAPTCHA image to make the characters inseparable by the OCR programs.*

vii. *Finally, apply a distortion to the generated CAPTCHA image by using some mathematical transformation. The application of the distortion makes characters within the image to twist and thus increases the difficulty for the OCR programs to decode it.*

Image-based CAPTCHA techniques involve the use of different patterns or concepts which the user needs to identify. Thus the algorithms for image generation, size and dimensions of generated CAPTCHA image, the size and types of the images in image database and the difficulty level of each vary significantly from one another. These techniques create a composite CAPTCHA image of required dimension and optionally add desired type of noise and other objects to it forming a complex background. Next, they select images or objects from the image database present at the server or by downloading the images from the Internet and apply various transformations like scaling, rotation, transparency, etc. to each and place them on the CAPTCHA image at desired positions. The resultant CAPTCHA image is shown to the user. The user needs to identify the object or concept presented in this image and act accordingly. The solution to CAPTCHA challenges may be a set of points on the CAPTCHA image or a text string or both. The solution to the CAPTCHA test is either stored in a database on the server along with other state information or in a cookie at the client computer. A Web service called CAPTCHA generation [23] is a new step towards generation of CAPTCHA schemes that provide CAPTCHA APIs which can be used by implicitly human interaction assumed Web services to defend against bots. They also provide code that permits programmers to generate customized CAPTCHA challenges.

## 4. Issues in CAPTCHA Design

### 4.1. Security Issues

Inefficiency of CAPTCHAs to resist to attacks aimed to break its underlying protocols via man-in-the middle or oracle attacks [24] due to advances in OCR techniques has necessitated designing of CAPTCHAs which are robust, secure and usable. Research works carried out in [5, 24, 25] discuss the inefficiency or report the breaking of various CAPTCHA techniques. Breaking a CAPTCHA means to solve a CAPTCHA challenge by writing a computer program [24]. It is a two step process vis-à-vis segmentation and character recognition [26]. Content Based Image Retrieval (CBIR) methods are used for segmentation of an image in regions, identification of regions of interest and extraction of semantic content expressed by the image or part of it. Success of an attack to break a CAPTCHA technique highly depends upon the accuracy obtained in the segmentation process Segmentation process chunks the image into letters and passes these chunks to the character recognition stage which attempts to map each chunk to a particular letter or digit. Jeff Yen *et al* in the research work reported in [24] demonstrate the breaking of most visual CAPTCHA schemes publicly available as Web service for CAPTCHA generation at captchaservice.org. A simple breaking procedure of text based CAPTCHAs involve segmentation of text string into individual segments, mapping of segments to characters and use of dictionary to guess the correct word. The complexity involved in segmentation process depends upon the complexity in algorithms used to generate the CAPTCHA test. A snake segmentation or geometric analysis may be used to further strengthen the segmentation process [24]. Character recognition may involve use of dictionary to facilitate the character recognition process in identifying the candidate text string. Segmentation process involves use of edge detection or thresholding to segment an image into regions. An edge detection technique detects outline of objects in an image by detecting jumps in its image intensity function. In simple thresholding pixels of image are set to white if their intensity exceeds a certain threshold value otherwise they are set to black. Background of an image can be separated from its foreground if threshold value(s) clearly separate(s) the two otherwise an adaptive thresholding which changes the intensity threshold for every pixel of the image in relation with the pixel intensity values of its neighboring pixels may be used to separate the foreground from the background. Shape matching [27], though a complex and time consuming process may be used to break an image CAPTCHA technique. It involves image collection of previously presented images, restoring these images to their original form and then comparing them with images presented in the subsequent tests to reveal difference between them. Random guessing wherein an attacker may click randomly on any portion(s) of the presented image to pass the test is yet another possible attack to defeat image based CAPTCHAs. Proxy through unaware users [22] also called laundry attack may be used by an attacker to break the CAPTCHA test. In this type of attack, attackers may download the CAPTCHA image and present it to unaware users on unrelated Web sites controlled by the attackers. As an example, an attacker may control a large network of pornographic or other similar Web sites where a visitor would be prompted with a CAPTCHA test to access a resource or download a torrent. The user not aware of the underlying mistrust solves the CAPTCHA challenge. Thus obtained response is sent by the Web site to the attacker who uses the received response along with other asked properties to gain access to the protected Web service. Further, a weak Implementation [28] of any CAPTCHA technique like allowing a session ID authorized by a single successful challenge to be re-used repeatedly to gain access to some protected service make that CAPTCHA technique insecure. CAPTCHA solutions that are stored in cookies of the client computers also make the test vulnerable to attacks.

### 4.2. Securing CAPTCHAs against attacks

A CAPTCHA test may be considered secure that is at least as expensive for a hacker as it would cost him using human operators [8, 28]. The security of a particular CAPTCHA test can be analyzed by investigating its resistance to attacks that possibly may be used to break it [28]. Further, tests against real users and bots can greatly help in ascertaining its security state.

Various methods are used to make text-based CAPTCHAs difficult to break. These methods include font tricks, choice of letters, noise, color model, overlap, distortion and degradation [26]. To make text CAPTCHAs secure against dictionary attacks, a complex background and some random object like circles, arcs, lines, etc. are added to the CAPTCHA image. This makes segmentation process difficult as it results in images of inter-connected components. The presences of complex background or clutter impose challenges to perform visual concept detection and identification [29] making CAPTCHAs secure against segmentation. Security against shape matching and segmentation can be improved by distorting CAPTCHA images by application of transformations like scaling, rotation and transparency. This makes restoration of images to original form difficult which is required for successful shape matching. The probability of a successful random guess can be decreased by increasing the area of CAPTCHA image and decreasing that of each sub image. Making a CAPTCHA image meaningful only in the specific context of the Web site that is protected will make CAPTCHA image not fungible and thus secure against laundry attacks [21]. Invalidating the CAPTCHA image after a specific time can also be used as a solution against laundry attacks [30]. Insecurity on account of weak implementation of CAPTCHA methods can be overcome by careful analysis, code reviews and timely updates. Use of encryption or hashing algorithm to secure CAPTCHA results either in cookies on client computers or database on the Web server minimizes security vulnerabilities of CAPTCHA. Further, CAPTCHA implementation should employ Global Unique Identifier (GUID) to ensure that sender of the CAPTCHA solution is really the computer which was send a CAPTCHA challenge by the server.

### 4.3. Usability Issues

Usability is a measure of the effectiveness, efficiency and satisfaction with which specified users can achieve specified goals in a particular environment [31]. Accuracy, response time and perceived difficulty of using a CAPTCHA scheme determine the usability of a particular CAPTCHA test. Accuracy is a measure of correctness with which users can respond to a CAPTCHA challenge without making mistakes. Response time is the time taken by a user to react to the CAPTCHA challenge. Perceived difficulty is the difficulty observed by the users in solving CAPTCHA challenges. High accuracy, low response time and low perceived difficultly are desired to make CAPTCHAs usable.

Distortion is used in CAPTCHAs to improve its security control; however, the use of excessive or unmanaged distortion level and methods may not only make CAPTCHAs unusable but also will lower its security control because the system would have to allow multiple attempts for failed tests [32]. Distortions also create ambiguous characters, hard to apart from each other and identify and thus are unfriendly to foreigners who are not native speakers of the language in which CAPTCHA is implemented. Inappropriate or unorganized CAPTCHA or any unsolicited or offensive image or text appearing in the CAPTCHA content will considerably reduce its usability. A CAPTCHA test can alienate or even frustrate a legitimate user if its presentation is poor. Use of color enhances the user interface but its misuse can cause both usability and security problems [32]. Research work carried out in [32] reported effective segmentation of overlapped characters generated through Cryptograph CAPTCHA by picking up pixels from the CAPTCHA image having same color. A CAPTCHA user interface may require a user to input response by typing characters in a text box or by selecting the answers from a dropdown list or by clicking on correct portion(s) of the CAPTCHA image. Most of the existing CAPTCHA challenges particularly text-based challenges use a text box to input response from the user, which in comparison to other user interfaces besides being difficult to work with is time consuming. An optimum size of CAPTCHA image is highly desired to have a balance between usability and security [28]. Large dimensions reduce the chances of successful blind attacks and thus improve security control. It also fastens the response time due to improved visibility of the sub-images or text within the CAPTCHA image. However, large images involve huge server processing for applying transformations and for transferring the images from the image database which decrease the performance. On the other side, smaller images offer advantages of faster image downloads and occupy less screen area making it easy to integrate the CAPTCHA challenge within the Web page. Further, use of CAPTCHAs has posed Web Accessibility challenge that could create a digital divide between normal and disabled users of Web Services.

### 4.4. Improving Usability

Jeff Yen and Ahmad Salar El Ahmad in their recent study [32] have provided a three dimensional framework for examining and improving the usability of CAPTCHAs. Under each dimension namely distortion, content and presentation, various usability issues have been identified and explained.

Distortion should be applied in a controlled manner to avoid creation of ambiguous characters or images which with some difficulty level can be identified. The identification of the sub-images or embedded characters can be made easy by showing current a portion of the CAPTCHA image in zoomed-in state when the user hovers over that portion of the image. Recently proposed Partial Credit Algorithm [28], in which "almost right" answers are treated as strong evidence that a user is human can be used to improve usability against complex distortions. The maximum and minimum allowable distortion levels for each sub-image may be automatically controlled by keeping track of images that legitimate users have failed to recognize. Use of well known images i.e. images which can be recognized easily by most of the users will improve usability of image-based CAPTCHAs. To improve the presentation of a CAPTCHA test various usability issues that must be addressed while designing a CAPTCHA challenge, are use of color, user interface and appropriate screen area so as to make the challenge simple, easy to answer, easy to integrate with the Web page and highly accurate [33]. A clickable interface simplifies and speeds-up the entry of the CAPTCHA solutions which improves user friendliness and permits the use of CAPTCHA on devices with small displays where they would otherwise be unusable. Using multiple types of CAPTCHAs like audio CAPTCHA along with text or image based CAPTCHA has become an alternative for visually impaired people to gain access to the protected Web resource. Further, guidelines that have been provided by World Wide Web consortium's (W3C) Web Accessibility Initiative (WAI) to make the Web sites accessible to the users with disabilities which must be followed to make them accessible to disabled persons. It is highly recommended to carry out usability tests of a CAPTCHA method against humans to ascertain its usability. The tests should analyze accuracy, response time and user satisfaction on different image sizes with different distortion levels.

### 5. Conclusion

Text-based CAPTCHA are most widely deployed and are in use since years in major web sites. Further, they are intuitive to users and can provide strong security if properly designed. Early text-based CAPTCHAs were straight forward for humans to solve. Advances in OCR techniques and consequently efficiency of bots in breaking text-based CAPTCHAs improved as a result of which text-based CAPTCHAs are designed harder. Currently text-based CAPTCHAs have become sufficiently hard for humans to solve and thus their usability has decreased at least for an ordinary user. Often ordinary users fail to solve hard text-based CAPTCHAs in their first attempt. Text-based CAPTCHA techniques have localization issues and thus are not friendly to foreigners. They use text box to input responses from the users, which in comparison to other user interfaces besides being difficult to work with, is also time consuming. Image-based CAPTCHA schemes have been proposed as an alternative to text-based CAPTCHAs but they have not been able to replace text-based. The Web page area required for displaying the CAPTCHA image and size of CAPTCHA image in all image-based CAPTCHA schemes is more in comparison to that required in text based methods. Further, the processes involved in creating image database, its storage requirement at the server and the delay caused by image processing at the server with each page refresh limits the use of image-based CAPTCHA schemes. Breaking a CAPTCHA challenge is difficult and it is very rare to find 100% success rate, however several CAPTCHA implementations have been broken and thus are proved to be inefficient. The advances in OCR techniques in terms of pattern recognition and computer vision have made CAPTCHAs prone to more and more attacks and thus, it is imperative to validate the robustness and effective usability of new CAPTCHA technique rigorously. A huge scope for research exists in designing new and novel CAPTCHA techniques that are user friendly, require less server processing and offer improved security control against bots.

## Biographies

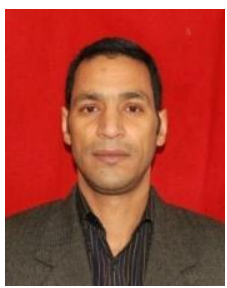

**M. Tariq Banday** was born in 1969. He did his M. Sc. and M. Phil. Degrees from the Department of Electronics, University of Kashmir, Srinagar, India in 1996 and 2008 respectively. He did advanced diploma course in computers and qualified UGC NET examination in 1997 and 1998. At present he is working as Assistant Professor in the Department of Electronics & Instrumentation Technology, University of Kashmir, Srinagar, India. He has to his credit several research publications in reputed journals and conference proceedings. He is a lifetime member of Computer Society of India and International Association of Engineers. His current research interests include Network Security, Internet Protocols and Network Architecture.

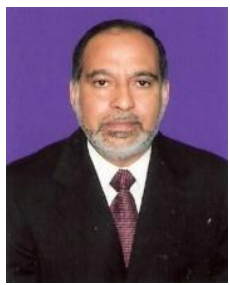

**Nisar A. Shah** was born in 1953. He did his M. Sc. and Ph. D. Degrees from the department of Physics, University of Kashmir, Srinagar, India in 1976 and 1981 respectively. At present he is working as Professor in the Department of Electronics & Instrumentation Technology, University of Kashmir. He has to his credit about 150 research publications which have been published in national and international journals of repute. He has supervised several research scholars in M. Phil. and Ph. D. programs. His current research interests include Digital Signal Processing and Network Security.